\documentstyle[12pt]{article}
\begin{document}
\newcommand{\be}{\begin{equation}}
\newcommand{\ee}{\end{equation}}
\newcommand{\bea}{\begin{eqnarray}}
\newcommand{\eea}{\end{eqnarray}}
\def\pd{\partial}
\def\no{\nonumber}
\baselineskip=24pt plus 2pt

\vspace{2mm}
\begin{center}
{\large \bf Master-Slave Scheme and Controlling Chaos 
in the Braiman and Goldhirsch Method} \\
\vspace{10mm}
Han-Tzong Su\footnote{E-mail address: htsu@mail.ncku.edu.tw}, 
Yaw-Hwang Chen\footnote{E-mail address:    
yhchen@ibm65.phys.ncku.edu.tw}, 
and Rue-Ron Hsu\footnote{E-mail address: 
rrhsu@mail.ncku.edu.tw}\\
\vspace{6mm}
Nonlinear Science Group, Department of Physics \\
National Cheng Kung University\\
Tainan,Taiwan 70101, Republic of China\\
\end{center}
\vspace{5mm}
\begin{center}
{\bf ABSTRACT}
\end{center} 
~~~~This brief report presents a master-slave 
scheme to demonstrate how control chaos works in the 
Braiman and Goldhirsch method for the one-dimensional map system.  
The scheme also naturally explain why the anomalous response arises in a periodically perturbed, unimodal map system. 
\\
PACS number(s): 05.45.+b.

\hfill{Typeset using \LaTeX}
\newpage 
~~~~ 
 Braiman and Goldhirsch (BG)~\cite{1} proposed a 
simple non-feedback method, in contrast to the Ott-Grebogi-Yorke feedback scheme~\cite{2}, to create stable periodic orbits 
from a chaos using a weak periodic perturbation.  Though there 
are some successful numerical and experimental demonstrations 
of the BG method~\cite{3}-\cite{5}, the periodicity and the stability condition of the targeted stable state was not identified analytically until recently~\cite{6}.  If one considers a 
generic one-dimensional chaotic map under the influence of a 
periodic perturbation, 
\be
z_{n+1}=f(z_n)-\alpha y_n,
\ee
where $\alpha$ is a small number and $y_n$ is the added weak 
 perturbation with periodicity $p$.  The desired stable states in response to the 
period-$p$ perturbation in a chaotic map system can only have 
the periodicity $q=kp$, where $k$ is an integer number. 
Furthermore, using the linear stability analysis, one can deduce 
that the stability condition, for the output with the periodicity 
$q=kp$, is 
\be
\vert M \vert =\Bigg\vert\prod_{j=1}^k\left[\prod_{\ell=1}^p\left(\left. \frac {\pd f}{\pd z}
\right\vert_{\bar z_{j\ell}} \right)\right]\Bigg\vert <1.
\ee
Here, $\bar z_{j\ell}$ is the ${j\ell}$ times mapping of $\bar 
z_{1}$, and $\bar z_{1}$ are the roots of the periodicity condition 
\be
z=f(f(...(f(f(z)-\alpha y_1) - \alpha y_{2})-...) -\alpha y_{kp-
1})- \alpha y_{kp}. 
\ee

Even though, the analysis is presented in an elegant 
mathematical form in the reference [6]. Providing a 
more intuitive picture to illustrate how a chaotic system can be 
controlled by weak periodic perturbation is still a worthwhile effort.  In this brief report, we  will 
introduce a conceptual picture, called the master-salve 
scheme, to explain how control chaos works in the 
BG method.  This picture will also gives us a new handle to 
understand the anomalous responses in a dynamical system 
under the influence of periodic perturbation.  For example, when a period-2 perturbation with elements $\{y_1= a, ~y_2=0.2\}$ is added to a chaotic logistic map, then the system becomes $z_{n+1}=4z_n(1-z_n)-y_n$.  
The antimonotonicity - concurrent creation and destruction of periodic orbits~\cite{7,8}, appears in the bifurcation diagram for the variation of $a$, see Fig.1.  It seems to be in the
 contrary to the well known numerical fact~\cite{9}: 
{\sl the antimonotonicity could never appear in a unperturbed one-dimensional 
unimodal map system}. In the next paragraph, we will see that this 
anomalous response can be interpreted naturally in 
the master-slave scheme.

To begin with, let us consider a generic map under the influence of a  period-$p$ orbit $\{  y_1, y_2, ..., y_{p}\}$, see Eq.(1) . For 
convenience, we will label the initial data and initial 
perturbation as $z_1$ and $y_1$, respectively. The key idea of 
the master-slave scheme is as following. We divide the original dynamical 
variables $z_n$ into $p$ new variables, called $\{ x^{(1)}_m, x^{(2)}_m, ..., 
x^{(p)}_m\}$.  The relation between $z_n$ and the new 
variables $x^{(i)}_m$ is defined as 
\be
x^{(i)}_m=z_{pm+i}, ~~~1\le i \le p.
\ee
Hence, the original dynamical equation can be separated 
into $p$ maps:
\bea
x^{(2)}_m&=&f(x^{(1)}_m)-\alpha y_1,\no\\
x^{(3)}_m&=&f(x^{(2)}_m)-\alpha y_2,\no\\
&...&\no\\
x^{(p)}_{m}&=&f(x^{(p-1)}_m)-\alpha y_{p-1},\no\\
x^{(1)}_{m+1}&=&f(x^{(p)}_m)-\alpha y_{p}.
\eea
Plugging the first $(p-1)$ maps, $x^{(2)}_m$,$x^{(3)}_m$,...$x^{(p)}_m$, into 
 $x^{(1)}_{m+1}$, one find the map between 
$x^{(1)}_{m+1}$ and $x^{(1)}_{m}$, which characterise the 
dynamical properties of original system, and let us call it {\sl the master equation}:
\bea
x^{(1)}_{m+1}&=&F(x^{(1)}_{m}; \alpha y_1, \alpha y_2,...,\alpha 
y_{p}) \no\\
&=&f(f(...(f(f(x^{(1)}_m)-\alpha y_1) - \alpha y_2)-...) -\alpha 
y_{p-1})- \alpha y_{p},  
\eea
The remained $p-1$ maps, which just are mappings 
of $x^{(1)}_{m}$, and are designated as {\sl the slave equations}:
\bea
x^{(2)}_m&=&f(x^{(1)}_m)-\alpha y_1,\no\\
x^{(3)}_m&=&f(x^{(2)}_m)-\alpha y_2,\no\\
&...&\no\\
x^{(p)}_{m}&=&f(x^{(p-1)}_m)-\alpha y_{p-1}.
\eea
Since, the dynamics of the slave equations are completely controlled by the master equation, hence the name-{\sl the master-slave scheme}.  As long as the master equation, Eq.(6), is in a stable 
period-$k$ orbit, then the slave equations indicate that $(p-1)$ 
images will appear simultaneously. It means that there exists a 
period-$kp$ orbit in the original system.

From linear stability analysis, one can deduce that the stability 
condition for the period-$k$ orbit in the master equation is $\vert 
M \vert < 1$.  The stability quantity M now simply is 
\be
M= \Bigl[\frac {\pd}{\pd x} F^k(x^{(1)}_{m}; \alpha y_1, \alpha 
y_2,...,\alpha y_{p}) \Bigr]
\Big\vert_{\bar x^{(1)}},
\ee
where $\bar x^{(1)}$ is one of the roots of the periodicity condition 
\be
\bar x^{(1)}=F^k(\bar x^{(1)}_{m}; \alpha y_1, \alpha y_2,...,\alpha 
y_{p}). 
\ee
Obviously, in terms of the original dynamical variable 
$z_n$ and the map $f$, Eq.(8) and Eq.(9) will reduce to 
Eq.(2) and Eq.(3), respectively.

Now, to be more specific, let us take $\alpha=1$ and the perturbation 
$y_n$ be of period-2 with elements $\{y_1= a, ~y_2=0.2\}$.  We 
will further assume that the system is a chaotic logistic map, $f(z)=4z(1-z)$, 
before we turn on the perturbation.  In this special case, the master 
equation becomes
\be
x^{(1)}_{m+1}=4(4x^{(1)}_{m}(1-x^{(1)}_{m})-a)(1-4x^{(1)}_{m}(1-
x^{(1)}_{m})+a)-0.2,
\ee
and the slave map is 
\be
x^{(2)}_{m}=4x^{(1)}_{m}(1-x^{(1)}_{m})-a.
\ee
Here, $x^{(1)}_{m}$ ($x^{(2)}_{m}$) denotes the odd (even) part of 
$z_m$, i.e. $x^{(1)}_{m}=z_{2m+1}$ ($x^{(2)}_{m}=z_{2m+2}$), and 
the initial value is labelled as $x^{(1)}_{0}=z_1$.  The 
bifurcation diagram of the master equation, Eq.(10), for the perturbation $a$ with values between 0.0 and 0.55, is shown in Fig.2.  The bifurcation diagram indicates that the desired stable 
period-$k$ orbit will appear if a suitable perturbation 
is applied on this chaotic logistic map.  For example, period-1 orbit 
occurs when $a$ is between $(0.194, 0.240)$; and period-2 orbits can be generated when $a$ is at $(0.170, 0.194)$, $(0.240, 0.290)$, or $(0.425, 0.464)$; etc.   
From Eq.(11), the bifurcation diagram the slave map is plotted in Fig.3.  One can see that the same periodic 
orbits also appear exactly at the same regions of the perturbation $a$.  Obviously, the 
combination of Fig.2 and Fig.3 leads to Fig.1 exactly.  Also, from the numerical simulation presented in the Fig.1, one can clearly see that the periodicity of 
the stable response in the chaotic map under the influence of a period-2 
perturbation is $2k$ - as one would have expected.

For a careful reader, she/he may have noticed that  there are some 
anomalous responses in the bifurcation diagrams of this perturbed logistic map, which occurs at 
the region $a\in (0.25, 0.45)$ , see Fig. 1-3. These 
anomalous responses are called the antimonotonicity that does not exist in a unperturbed logistic map.  Since the logistic map $f(z)=rx(1-x)$ 
only has one critical point at $x=0.5$, it seems to be a dilemma for those 
who are familiar with the work of Dawson, Grebogi and 
Ko\c{c}ak~\cite{9} : {\sl if a one-dimensional map 
$x_{n+1}=F(x_n,\alpha)$ has at least two critical points that lie in a 
chaotic attractor for a parameter $\alpha=\alpha^*$, then 
generically, F is antimonotone at $\alpha^*$.}  However, as it has been 
mentioned in the last paragraph, the dynamical of a periodically 
perturbed system is governed by the master equation.  The 
right hand side of Eq.(10) is a fourth order polynomial of $x^{(1)}_{m}$, and this
implies that the system may have three critical points.  Therefore, 
the antimonotonicity could arise naturally in a periodically 
perturbed logistic map.

Finally, a brief note is given as a concluding remark. The master-slave scheme gives us a conceptual picture that 
the periodic perturbation indeed makes controlling chaos feasible.  The scheme also helps us to understand how the anomalous responses arise in a periodically perturbed one-dimensional map.

\vspace{5mm}

\begin{flushleft}
{\Large\bf Acknowledgements}\\
\end{flushleft}
~~~~This work is supported in part by the National Science Council, 
Taiwan R.O.C. under the contact numbers NSC 87-2112-M006-012 and NSC 87-2811-M006-007.

\newpage
\begin{flushleft}
{\Large\bf Figure Captions :}\\
\end{flushleft}
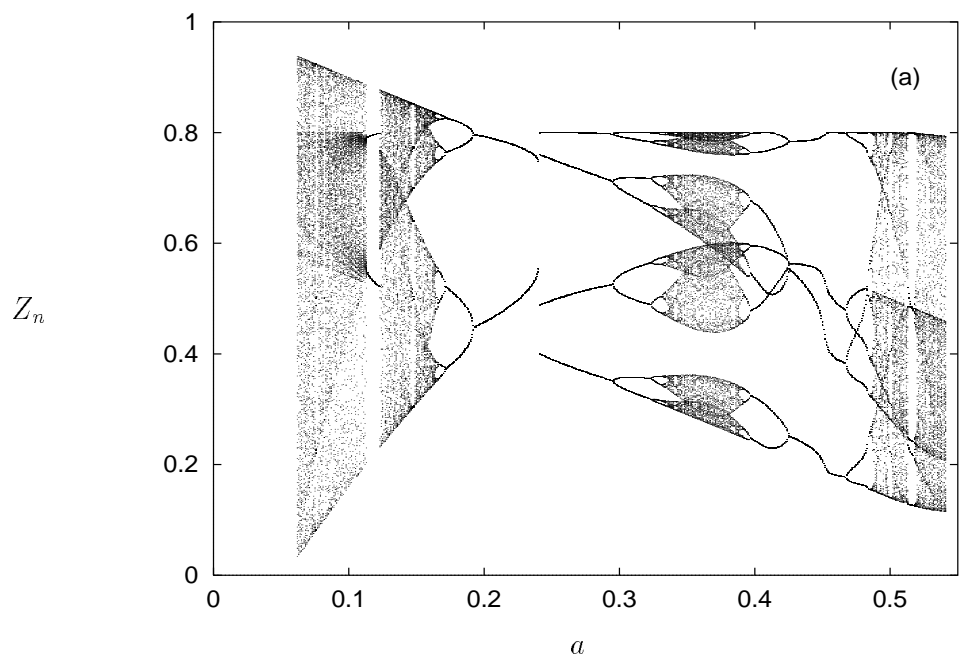
{\bf Figure 1.}
Bifurcation diagram for a periodically perturbed chaotic logistic map, 
$z_{n+1}=4z_n(1-z_n)-y_n$, where $y_n$ is of period-2 and with elements $\{y_1= a, ~y_2=0.2\}$. 
The  initial point for $z_1$ is $0.54$, and 100 data points are plotted after 4000 transient iterations. 
 \\
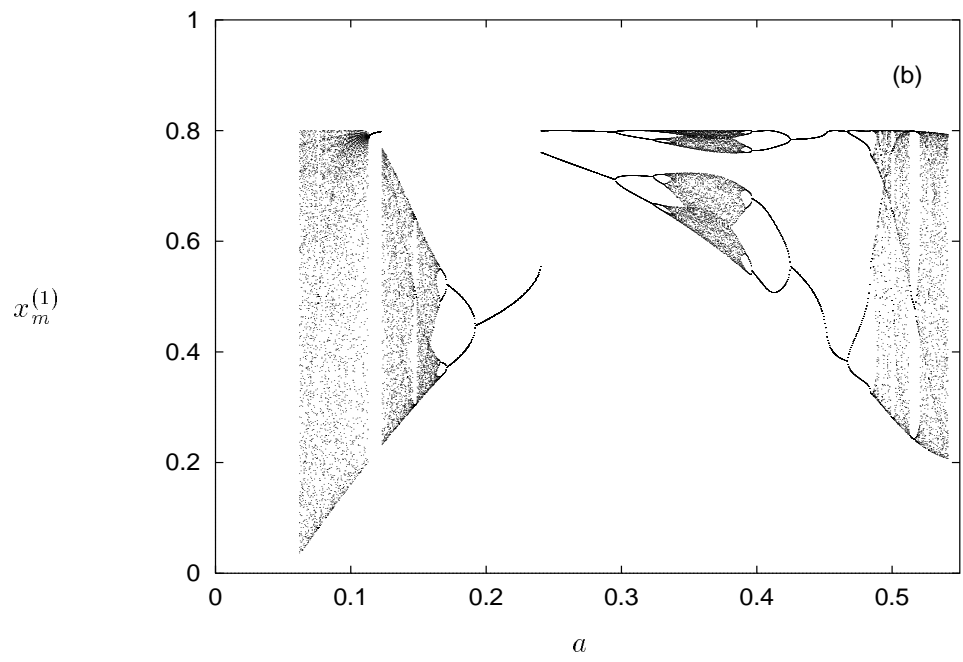
{\bf Figure 2.}  
Bifurcation diagram for the master equation Eq.(10), versus the perturbation $a$. The initial point for $x^{(1)}_0$ is $0.54$, and  50 data points are plotted after 2000 transient iterations. \\
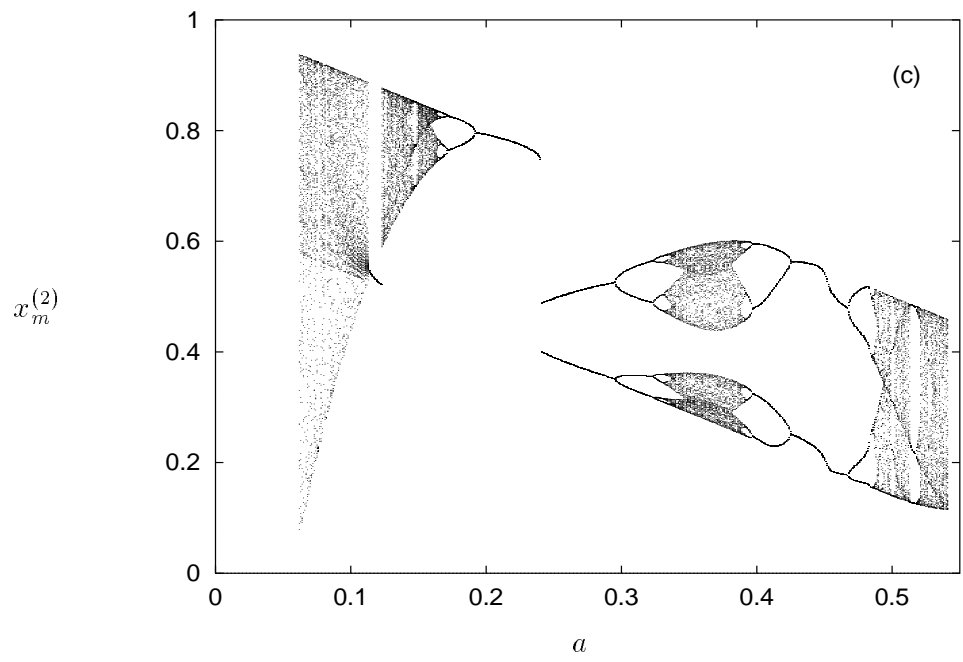
{\bf Figure 3.}  
The image $x^{(2)}_m$, which is determined by the slave equation Eq.(11), of the master equation Eq.(10) verus $a$.\
\end{document}